\documentclass[aps,pre,12pt,a4paper,preprint,amsmath,amssymb,groupedaddress]{revtex4-1} 

\usepackage{graphicx}

\usepackage{times}
\usepackage{mathptmx}

\usepackage{bm}
\usepackage{amsmath}

\usepackage{color}

\newcommand{\pdf}{{PDF}}
\newcommand{\sol}{{SOL}}
\newcommand{\kstar}{{KSTAR}}

\newcommand{\rmd}{\text{d}}

\newcommand{\rme}{\text{e}}
\newcommand{\rmG}{\text{G}}

\newcommand{\rmw}{\text{w}}

\newcommand{\rmcm}{\text{cm}}

\newcommand{\taud}{\tau_\rmd}
\newcommand{\tauw}{\tau_\rmw}
\newcommand{\Jrms}{J_\text{rms}}

\newcommand{\nebar}{\ensuremath{\overline{n}_{\rme}}}
\newcommand{\nG}{n_\rmG}
\newcommand{\Cs}{\ensuremath{C_{\mathrm{s}}}}

\newcommand{\Phiave}{\ensuremath{\langle{\Phi}\rangle}}
\newcommand{\Phirms}{\ensuremath{\Phi_\mathrm{rms}}}

\newcommand{\wt}{\widetilde}

\newcommand{\ave}[1]{{\left<#1\right>}}

\newcommand{\Refs}[1]{Refs.~\onlinecite{#1}}
\newcommand{\Eqref}[1]{Eq.~\eqref{#1}}

\newcommand{\Figref}[1]{Fig.~\ref{#1}}

\newcommand{\Secref}[1]{Sec.~\ref{#1}}

\newcommand{\Tabref}[1]{Tab.~\ref{#1}}

\newcommand{\CPP}{\textit{Contrib.\ Plasma Phys.}}
\newcommand{\JGR}{\textit{J.~Geophys.\ Res.}}
\newcommand{\JNM}{\textit{J.~Nuclear Mater.}}

\newcommand{\NF}{\textit{Nucl.\ Fusion}}

\newcommand{\PF}{\textit{Phys.\ Fluids}}
\newcommand{\PFB}{\textit{Phys.\ Fluids~B}}
\newcommand{\PPCF}{\textit{Plasma Phys.\ Contr.\ Fusion}}
\newcommand{\PFR}{\textit{Plasma Fusion Res.}}

\newcommand{\PP}{\textit{Phys.\ Plasmas}}
\newcommand{\PRE}{\textit{Phys.\ Rev.~E}}

\newcommand{\PRL}{\textit{Phys.~Rev.\ Lett.}}

\begin{document}

\title{SOL width and intermittent fluctuations in KSTAR}

\author{O.~E.~Garcia}
\email{odd.erik.garcia@uit.no}
\author{R.~Kube}
\author{A.~Theodorsen}
\affiliation{Department of Physics and Technology, UiT The Arctic University of Norway, N-9037 Troms{\o}, Norway}

\author{J.-G.~Bak}
\author{S.-H.~Hong}
\author{H.-S.~Kim}
\author{the KSTAR {P}roject {T}eam}
\affiliation{National Fusion Research Institute, Daejeon, Republic of Korea}

\author{R.~A.~Pitts}
\affiliation{ITER Organization, CS 90 046, 13067 St Paul Lez Durance Cedex, France}

\begin{abstract}
Radial profiles of the ion saturation current and its fluctuation statistics are presented from probe measurements in L-mode, neutral beam heated plasmas at the outboard mid-plane region of KSTAR. The familiar two-layer structure, seen elsewhere in tokamak L-mode discharges, with a steep near-\sol\ profile and a broad far-\sol\ profile, is observed. The profile scale length in the far-\sol\ increases drastically with line-averaged density, thereby enhancing plasma interactions with the main chamber walls. Time series from the far-\sol\ region are characterised by large-amplitude bursts attributed to the radial motion of blob-like plasma filaments. Analysis of a data time series of several seconds duration under stationary plasma conditions reveals the statistical properties of these fluctuations, including the rate of level crossings and the average duration of periods spent above a given threshold level. This is shown to be in excellent agreement with predictions of a stochastic model, giving novel predictions of plasma--wall interactions due to transient transport events.
\end{abstract}

\date{\today}

\maketitle
\clearpage

\section{Introduction}

The boundary region of magnetically confined plasmas is generally in an inherently fluctuating state. Single point measurements of the plasma density reveal frequent occurrence of large-amplitude bursts and relative fluctuation levels of order unity \cite{dmz,garcia-pfr}. These fluctuations, seen in the scrape-off layer (\sol) of all tokamaks, are attributed to radial motion of blob-like filamentary structures through the \sol, leading to broad profiles and enhanced levels of plasma interactions with the main chamber walls that may be an issue for next generation magnetic confinement experiments \cite{pitts1,pitts2,whyte,lipschultz-nf,labombard2,labombard1,asakura,lipschultz-ppcf}.

Measurements from a number of tokamak experiments have demonstrated that as the line-averaged plasma density increases, the radial particle density profile in the \sol\ becomes broader and plasma--wall interactions increase \cite{asakura,lipschultz-ppcf,whyte,labombard1,labombard2,rudakov-nf,garcia-tcv-jnm,garcia-tcv-nf,garcia-tcv-ppcf2,labombard3,carralero-nf,carralero-jnm,militello-nf,carralero-prl}. The particle density profile typically exhibits a two-layer structure. Close to the separatrix, in the so-called near-\sol, it has a steep exponential decay and moderate fluctuation levels. Beyond this region, in the so-called far-\sol, the profile has an exponential decay with a much longer scale length and a fluctuation level of order unity \cite{asakura,lipschultz-ppcf,labombard1,labombard2,rudakov-nf,garcia-tcv-jnm,garcia-tcv-nf}. As the discharge density limit is approached, the profile in the far-\sol\ becomes broader and the break point moves radially inwards such that the far-\sol\ profile effectively extends all the way to the magnetic separatrix or even inside it \cite{rudakov-nf,garcia-tcv-jnm,garcia-tcv-nf,garcia-tcv-ppcf2,labombard3}.

The first part of this contribution augments the tokamak \sol\ profile database by presenting in \Secref{sec:profiles} a summary of the first \sol\ profile measurements on the Korean Superconducting Tokamak Advanced Research (\kstar), obtained at the outboard mid-plane of lower single null diverted, L-mode discharges \cite{bak1,bak2}. The results presented here are consistent with measurements on many other devices, in particular the increase of profile scale length with increasing line-averaged density. Moreover, the relative fluctuation level and the skewness and flatness moments are shown to vary weakly with radial position and line-averaged density in the far-\sol, suggesting the same kind of robustness of fluctuations found for many other devices \cite{labombard1,labombard2,rudakov-nf,garcia-tcv-jnm,garcia-tcv-nf,garcia-tcv-ppcf2,labombard3}.

A novel stochastic model has been proposed in order to describe intermittent fluctuations in the \sol, based on a super-position of uncorrelated pulses with an exponential pulse shape of constant duration and exponentially distributed pulse amplitudes \cite{garcia-acm-jnm,garcia-acm-php,kube-acm-ppcf,garcia-tcv-nfl,theodorsen-tcv-ppcf,garcia-prl,kube-mse,garcia-php,theodorsen-php,theodorsen-pre}. Under some simplifying assumptions, this model predicts an exponential radial profile and thus elucidates the physical mechanisms underlying broad radial profiles and large fluctuation levels in the \sol\ \cite{garcia-php}. The stochastic model and its predictions are presented in \Secref{sec:model}.

To contribute further to the understanding of the statistical properties of plasma fluctuations in the \sol\ and to contribute to the cross-machine scaling of this turbulence, a dedicated experiment was performed on \kstar\ with a reciprocating Langmuir  dwelled at a fixed position in the far-\sol\ during an entire discharge. This yielded high frequency turbulence measurements over a period of 5.5 seconds, several factors longer than previously obtained on other tokamaks. The resulting ion saturation current time series of unprecedented duration is analysed and compared with a similar investigation of a realisation of the stochastic model with additional noise in \Secref{sec:burst}. Excellent agreement is found between the two time series, including large-amplitude burst events and an analysis of level crossings and the average duration of time intervals spent above a given threshold level \cite{garcia-php,theodorsen-php,theodorsen-pre,sato,fattorini}.

A discussion of the results, the conclusions and an outlook are given in \Secref{sec:discuss}. The \kstar\ measurements presented here give further evidence for universality of fluctuations in the boundary region of magnetically confined plasmas. These are here shown to be described by a novel stochastic model. This includes the rate of level crossing and excess times, which are crucial for threshold phenomena like plasma--wall interactions. The stochastic model thus has the potential to provide all relevant distributions as far as the pulse duration and the lowest order moments can be reliably predicted for fusion plasmas.

\section{Experimental setup}\label{sec:exp}

\begin{figure}
\includegraphics[width=7.5cm]{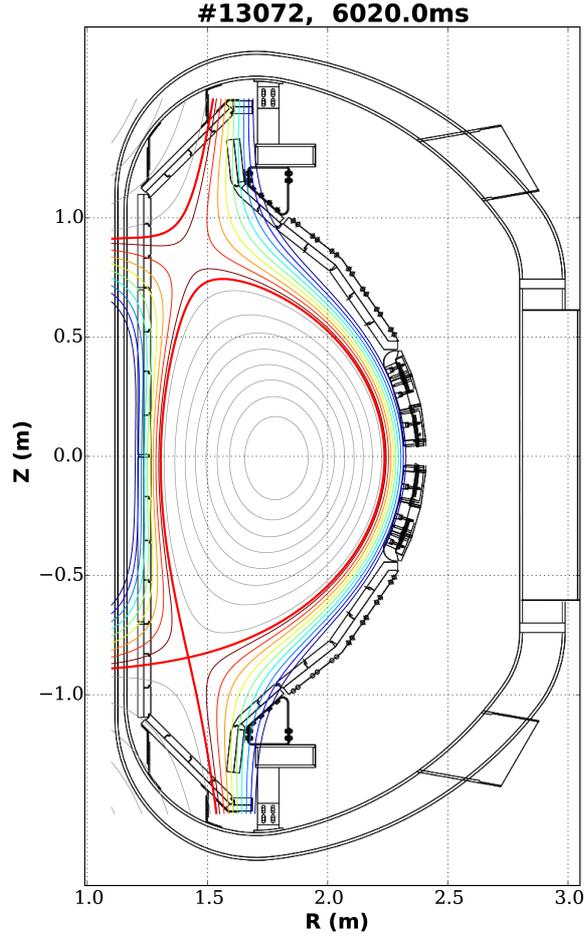}
\caption{Poloidal cross-section of \kstar\ with magnetic flux surfaces calculated from the magnetic equilibrium reconstruction of shot 13072. The reciprocating probe enters the \sol\ at the outboard mid-plane.}
\label{fig:kstar}
\end{figure}

Results are presented from reciprocating Langmuir probe measurements in lower single null, deuterium fuelled L-mode plasmas in \kstar\ \cite{bak1,bak2}. This superconducting, full carbon wall tokamak has a minor radius of $0.5\,$m and a major radius of $1.9\,$m. The experimental measurements were made with a plasma current of $0.6\,$MA, axial toroidal magnetic field of $2\,$T, neutral beam heating power of $1\,$MW and electron cyclotron resonance heating of $0.3\,$MW. For these parameters, the disruptive density limit is at $\nebar/\nG\approx0.6$, corresponding to complete divertor detachment, where $\nebar$ is the line-averaged density and $\nG$ is the Greenwald density \cite{greenwald}. A poloidal cross-section of \kstar\ is presented in \Figref{fig:kstar}, which also shows magnetic flux surfaces based on an equilibrium reconstruction for one of the discharges analysed in the following.

A fast reciprocating Langmuir probe assembly moves through the \sol\ region at the outboard mid-plane, measuring the ion saturation current with a sampling rate of $2\,$MHz. Only probe data from stationary phases of the plasma discharges are analysed, which typically have a duration of $8\,$s. Measurements influenced by probe arcing have been carefully eliminated from the probe data analysed here. A scan in line-averaged density up to the disruptive limit has been performed. Table~\ref{tab:symbols} gives the \kstar\ shot number, the Greenwald fraction of the line-averaged density and the plot marker and color used for the following presentation of the results. Further information about the probe system can be found in \Refs{bak1,bak2}.

\begin{table}[h]
\begin{center}
\begin{tabular}{ccc}

 \quad Shot number \quad & \quad $\nebar/\nG$ \quad & \quad Plot marker \quad \\

 \hline\hline

 13094 & 0.55 & {\textcolor{red}{\begin{math}{\blacktriangle}\end{math}}} \\

 13097 & 0.44 & {\scriptsize\textcolor{green}{\begin{math}\blacksquare\end{math}}} \\

 13095 & 0.34 & {\Large\textcolor{blue}{\begin{math}\bullet\end{math}}} \\

 13092 & 0.25 & \rotatebox{45}{\scriptsize\textcolor{magenta}{\begin{math}\blacksquare\end{math}}}\\

 13084 & 0.22 & {\textcolor{black}{\begin{math}\blacktriangledown\end{math}}} \\

 \hline

\end{tabular}
\end{center}
\caption{\kstar\ density scan experiments giving the shot number, Greenwald fraction of the line-averaged density, and the plot marker and color used in the following presentation of the results.}
\label{tab:symbols}
\end{table}

For each shot, the probe head moves through the outboard mid-plane \sol\ plasma up to a distance of $2.5\,\rmcm$ from the magnetic separatrix. Typically, two reciprocations are made per discharge, separated by several seconds. 
In the resulting time series of the ion saturation current, hysteresis is observed between the ingoing and outgoing profiles. This is likely due to perturbation of the plasma by the probe assembly. For this reason, only data for the inward probe motion and from one reciprocation for each plasma discharge is used for the following analysis. A parabolic function is fitted for the probe position versus time. Based on this, the data time series is divided into sub-records corresponding to $0.5\,\rmcm$ radial movement of the probe, giving of the order of $10^4$ or more data elements per bin. This has been found as the best compromise between spatial resolution and convergence of estimators for the lowest order statistical moments. From the resulting sub-records of typically $5\,$ms duration, the sample mean, standard deviation, skewness and flatness moments are readily calculated. The results from the density scan experiments are presented in \Secref{sec:profiles}.

\begin{figure}
\includegraphics[width=7.5cm]{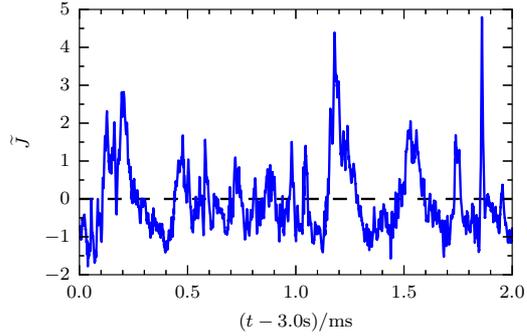}
\caption{Time series of the ion saturation current fluctuations showing frequent occurrence of large-amplitude bursts.}
\label{fig:jraw}
\end{figure}

In order to further investigate the statistical properties of large-amplitude fluctuations in the ion saturation current, a special experiment was performed with the probe maintained at a fixed position in the \sol\ throughout the entire discarge. The line-averaged density for this shot was $\nebar/\nG=0.3$, while all other parameters were the same as for the density scan described above (see \Figref{fig:kstar} for the magnetic equilibrium). The probe was placed $3.6\,\rmcm$ from the separatrix and $3.0\,\rmcm$ in front of the limiter structures. The resulting time series of the ion saturation current under stationary plasma conditions has an unprecedented duration of $5.5\,$s. A short part of this time series is presented in \Figref{fig:jraw}. Here and in the following, the rescaled ion saturation current signal is defined by $\wt{J}=(J-\overline{J})/J_\mathrm{rms}$, where $\overline{J}$ and $J_\text{rms}$ are the sample mean and root mean square values, respectively. The raw data presented in \Figref{fig:jraw} show the frequent occurence of large-amplitude bursts, which are typically observed in the boundary region of magnetically confined plasmas. In \Secref{sec:burst} the statistical properties of these fluctuations are investigated in detail.


\section{SOL profiles}\label{sec:profiles}

\begin{figure}
\includegraphics[width=7.5cm]{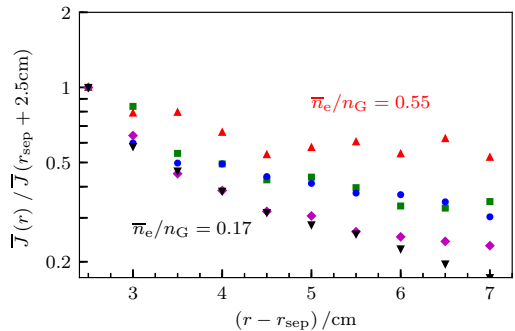}
\caption{Radial profiles of the ion saturation current for various line-averaged densities normalised to the value at the innermost position at $r=r_\text{sep}+2.5\,\text{cm}$. See \Tabref{tab:symbols} for the densities appropriate to each symbol.}
\label{fig:jave}
\end{figure}

The radial profiles of the ion saturation current are presented in \Figref{fig:jave} for the various line-averaged densities. A double-exponential function has been fitted to each profile, giving an estimate of the profile scale length in the near- and far-\sol\ regions and their variation with the line-averaged density. Each profile in \Figref{fig:jave} is normalized to the value of the profile at a distance of $2.5\,\rmcm$ from the estimated magnetic separatrix location. The familiar profile broadening with increasing line-averaged density is clearly observed. In the far-\sol, the scale length more than doubles from $3.4\,\rmcm$ at $\nebar/\nG=0.22$ to $8.6\,\rmcm$ at $\nebar/\nG=0.55$. At the highest line-averaged density, the ion saturation current profile is broad and well described by a single exponential function over the entire \sol\ measurement region, similar to what has been observed in many other experiments \cite{whyte,lipschultz-ppcf,asakura,labombard1,labombard2,garcia-tcv-jnm,garcia-tcv-nf,garcia-tcv-ppcf2,labombard3,rudakov-nf,carralero-nf,militello-nf,carralero-jnm,carralero-prl}.


\begin{figure}
\includegraphics[width=7.5cm]{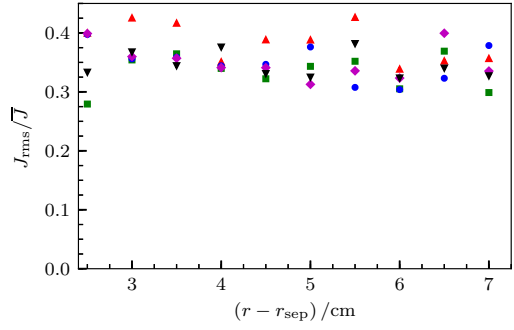}
\caption{Radial profiles of relative fluctuation level in the ion saturation current for various line-averaged particle densities.}
\label{fig:jrfl}
\end{figure}

\begin{figure}
\includegraphics[width=7.5cm]{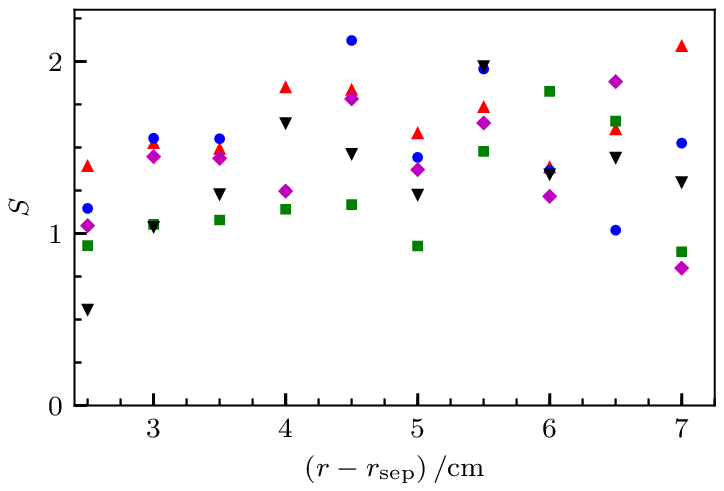}
\caption{Radial profiles of the sample skewness for the ion saturation current for various line-averaged particle densities.}
\label{fig:jskw}
\end{figure}

\begin{figure}
\includegraphics[width=7.5cm]{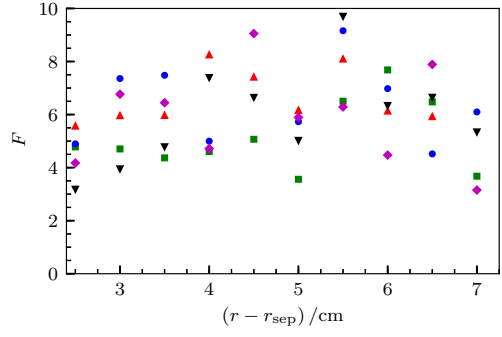}
\caption{Radial profiles of the sample flatness for the ion saturation current for various line-averaged particle densities.}
\label{fig:jflt}
\end{figure}

The radial profiles of the relative fluctuation level for the various line-averaged densities are presented in \Figref{fig:jrfl}. Here it is clearly seen that the fluctuation level lies at approximately 35\%\ throughout the \sol\ measurement region for all line-averaged densities investigated. The sample skewness moments presented in \Figref{fig:jskw} are larger than unity over most of the \sol. Similarly, the sample flatness moments in \Figref{fig:jflt} are significantly larger than three for most line-averaged densities and radial positions, which is the flatness value for a normally distributed random variable \cite{kube-mse,garcia-prl,garcia-php,theodorsen-php}. Due to the short duration of the time series, there is significant scatter of the data points for the higher order moments \cite{kube-mse,garcia-php}.

\begin{figure}
\includegraphics[width=7.5cm]{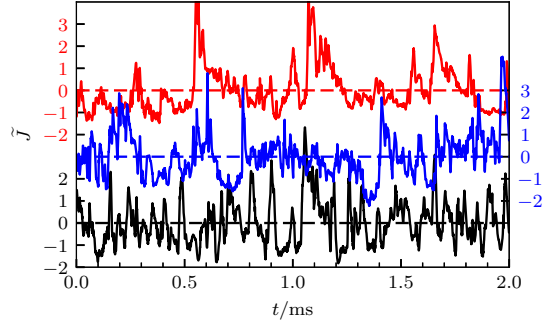}
\caption{Time series of the ion saturation current fluctuations at $2.5\,\text{cm}$ from the separatrix for the highest (top), intermediate (middle) and lowest (bottom) line-averaged densities.}
\label{fig:jraw-scan}
\end{figure}

It should, however, be noted that for the measurement point closest to the separatrix, the skewness and flatness moments are slightly larger for the highest line-averaged density, which has a broad profile across the entire \sol\ measurement region. This is consistent with the raw ion saturation current time series shown in \Figref{fig:jraw-scan} for the lowest, intermediate and highest line-averaged densities. Here it is clearly seen that the signal is more intermittent and dominated by large-amplitude bursts for the highest line-averaged density.

These results demonstrate that the plasma in the \sol\ of \kstar\ is in an inherently fluctuating state with positively skewed and flattened fluctuation amplitudes. Based on similar results from other devices, these fluctuations are attributed to the radial motion of blob-like plasma filaments. The mean profile becomes broader with increasing line-averaged density, thereby enhancing plasma interactions with the main chamber walls. The plasma--surface interactions depend on the rate of level crossings and the duration of intervals where the signal exceeds some threshold level. Before discussing these properties of the fluctuations, a stochastic model will first be described in the following section.

\section{Stochastic modelling}\label{sec:model}

Previous measurements in tokamak \sol\ plasma have shown that large-amplitude plasma fluctuations have on average an exponential wave-form with constant duration, exponentially distributed amplitudes, and appear in accordance with a Poisson process \cite{garcia-acm-jnm,garcia-acm-php,garcia-tcv-nfl,theodorsen-tcv-ppcf,kube-acm-ppcf}. This provides evidence for a stochastic model of the fluctuations as a super-position of uncorrelated pulses \cite{garcia-prl,garcia-php,theodorsen-php,kube-mse,theodorsen-pre},
\begin{equation} \label{fpp}
\Phi_K(t) = \sum_{k=1}^{K(T)} A_k\varphi(t-t_k) ,
\end{equation}
where $\varphi(t)$ is the pulse shape, $A_k$ is the amplitude and $t_k$ the arrival time for the pulse labeled $k$. It is assumed that the number of pulses $K(T)$ occurring during a time interval of duration $T$ is Poisson distributed and that the pulse arrival times $t_k$ are uniformly distributed on the interval $(0,T)$. From this it follows that the waiting times are exponentially distributed with the average waiting time given by $\tauw$ \cite{garcia-prl,garcia-php,theodorsen-php}. In the following, an exponential pulse shape will be considered,
\begin{equation}
\varphi(t) = \Theta\left(\frac{t}{\taud}\right)\exp\left(-\frac{t}{\taud}\right) ,
\end{equation}
where $\Theta$ is the unit step function and the pulse duration $\taud$ is taken to be the same for all pulses. For this stochastic process, the intermittency parameter $\gamma=\taud/\tauw$ determines the degree of pulse overlap and it can be shown that the probability density function (\pdf) approaches a normal distribution in the limit of large $\gamma$, independent of the amplitude distribution and pulse shape \cite{garcia-prl,garcia-php}.

\begin{figure}
\includegraphics[width=7.5cm]{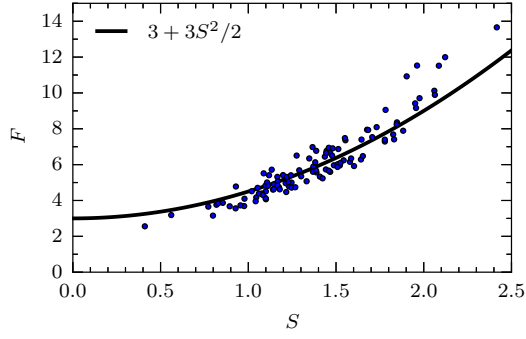}
\caption{Scatter plot of flatness versus skewness moments for the reciprocating probe data in the \kstar\ density scan.}
\label{fig:fvss}
\end{figure}

For the particular case of an exponential pulse shape and exponentially distributed pulse amplitudes, the stationary \pdf\ for the random variable $\Phi_K(t)$ is a Gamma distribution with the shape parameter given by $\gamma$ \cite{garcia-prl,garcia-php}. The mean value and variance of the signal are given by $\gamma\ave{A}$ and $\gamma\ave{A}^2$, respectively, where $\ave{A}$ is mean pulse amplitude, and there is a parabolic relationship between the skewness and flatness moments given by $F=3+3S^2/2$. A scatter plot of the sample flatness versus skewness moments for the \kstar\ density scan experiments discussed in the previous section is presented in \Figref{fig:fvss}. The parabolic relation is clearly a good description of these measurement data.

For exponential pulse shapes with duration $\taud$, the auto-correlation function for the random variable is readily calculated as $R_\Phi(\tau)=\ave{\Phi(t)\Phi(t-\tau)}=\Phiave^2+\Phirms^2\exp(-\tau/\taud)$. This allows the pulse duration time $\taud$ to be estimated for a synthetic data time series or experimental measurements. Furthermore, the stochastic model described above allows the rate of threshold crossings $X_\Phi(\Phi)$, the average duration $\langle\triangle T\rangle$ of time intervals where the process exceeds some prescribed threshold level, and how these change with the intermittency parameter $\gamma$, to be computed analytically \cite{garcia-php,theodorsen-php}.

Measurements of fluctuations in the \sol\ of tokamak plasmas have demonstrated that there is additional noise on top of the large-amplitude bursts that is not captured by the process given by \Eqref{fpp} \cite{garcia-acm-php,kube-acm-ppcf}. The effect of this additional noise can be described through a stochastic differential equation on the form \cite{theodorsen-pre}
\begin{equation} \label{sde}
\taud\,\frac{\rmd\Delta_K}{\rmd t} + \Delta_K = \sum_{k=1}^K A_k\delta\left( \frac{t-t_k}{\taud} \right) + \sigma\left( \frac{2}{\taud} \right)^{1/2}\xi(t) ,
\end{equation}
where $\xi(t)$ is a standard white noise process. The solution of this equation can be written as $\Delta_K=\Phi_K+\sigma Y$, where the Ornstein--Uhlenbeck process $Y(t)$ is normally distributed with vanishing mean and unit standard deviation. The process described by \Eqref{sde} has the same exponential decay response for the stochastic noise forcing $\xi(t)$ as for the Poisson point process $\Phi_K(t)$ described by \Eqref{fpp}. It should be noted that additional noise allows the signal to have negative values, as opposed to the process described by \Eqref{fpp}.

The auto-correlation function for the process $\Delta_K(t)$ is the same as for $\Phi_K(t)$, while the stationary \pdf\ for $\Delta_K$ is the convolution of a Gamma and a normal distribution \cite{theodorsen-pre}. Comparing this distribution to simulations of the process or experimental measurement data provides an estimate of the intermittency parameter $\gamma$ and the noise ratio $\epsilon=\sigma^2/\Phirms^2$ as fit parameters. This distribution has recently been shown to give an excellent description of plasma fluctuations in the \sol\ of Alcator C-Mod \cite{kube-acm-ppcf}. Closed analytical expressions for the level crossing rate and average excess times in the case of additional noise have not yet been derived.

\section{Fluctuation statistics}\label{sec:burst}

In this section, predictions of the stochastic model given by \Eqref{sde} are compared with probe measurements on \kstar. In order to critically assess the underlying assumptions and predictions of the model, a simulation of the stochastic process has been calculated using model parameters estimated from the long data time series from the probe dwell experiment discussed in \Secref{sec:exp}. Results are presented from an identical analysis of the measurement and synthetic data time series. In the following plots, a full blue line represents results from analysis of the \kstar\ ion saturation current time series, a dotted black line is the result of a similar analysis of the synthetic data, and a dashed green line is the best fit of using the analytical function specified in the following.

\begin{figure}
\includegraphics[width=7.5cm]{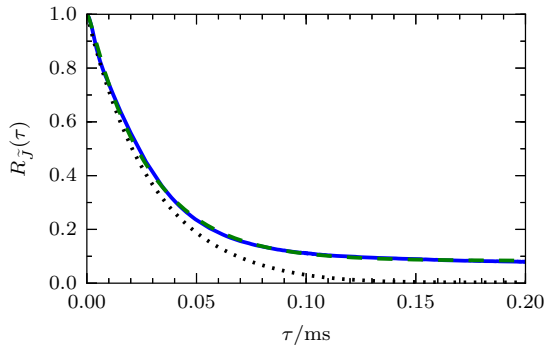}
\caption{Auto-correlation function for the ion saturation current (full blue line), the synthetic data (dotted black line) and the best fit of a modified exponential function  to the measurement data (dashed green line).}
\label{fig:jacf}
\end{figure}

The ion saturation current signal shown in \Figref{fig:jraw} is clearly dominated by the frequent appearance of large-amplitude bursts, which are generally characterised by an asymmetric wave-form with a fast rise and slower decay. It should be noted that the peak amplitudes of the ion saturation current bursts are typically several times the rms value.
The auto-correlation function for the ion saturation current signal and the synthetic data are presented in \Figref{fig:jacf}. The latter has the exponential decay predicted by the model. However, the auto-correlation function for the measurement data does not decay to zero and is in \Figref{fig:jacf} fitted by the modified exponential function $R_{\wt{J}}(\tau)=C+(1-C)R_{\wt{\Phi}}(\tau)$, where $R_{\wt{\Phi}}(\tau)=\exp(-\tau/\taud)$. This is clearly an excellent fit to the data and gives a correlation time of $\taud=30\,\mu$s, which is used as an input parameter for the model simulation.

\begin{figure}
\includegraphics[width=7.5cm]{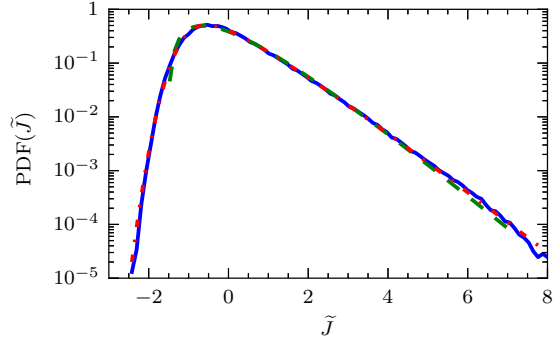}
\caption{Probability density function for the ion saturation current (full blue line), the best fit of a Gamma distribution (green dashed line) and the best fit of the convolution of a Gamma and a normal distribution (dash--dotted red line).}
\label{fig:jpdf}
\end{figure}

The \pdf\ for the ion saturation current signal is presented in \Figref{fig:jpdf}. Also shown in this figure is the best fit of a Gamma distribution, giving $\gamma=2.4$, and the best fit of the prediction of the stochastic model with additional noise, which is a convolution of a Gamma and a normal distribution. The latter provides an estimate for the model parameters, which in this case are given by $\gamma=1.7$ and $\epsilon=0.11$. The sample skewness and flatness moments for the ion saturation current time series are $1.3$ and $6.1$, respectively, in agreement with expectations from the stochastic model, which give $1.3$ and $5.9$, respectively. These values are also consistent with those found for the density scan experiments reported in \Secref{sec:profiles}. It should be noted that the distribution function covers more than four decades in probability, which is a result of the long data time series available here.

The saturation current \pdf\ is positively skewed and flattened and has an exponential tail towards large values, reflecting the frequent appearance of large-amplitude bursts in the time series. In order to reveal the statistical properties of these fluctuations,
a standard conditional averaging technique is utilised \cite{johnsen,huld,nielsen}. Events when the ion saturation current is above a specified amplitude threshold value are recorded. The algorithm searches the signal for the largest amplitude events, and records conditional sub-records centred around the time of peak amplitude whenever the amplitude condition is satisfied. These sub-records are then averaged over all events to give conditionally averaged wave-forms associated with large-amplitude events in the signal. Overlap of conditional sub-records are avoided in order to ensure statistical independence of the events.

\begin{figure}
\includegraphics[width=7.5cm]{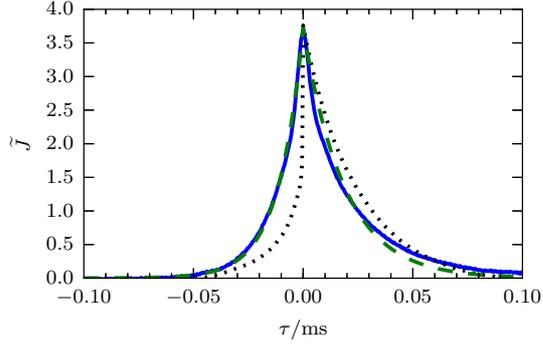}
\caption{Conditionally averaged wave-forms with peak amplitudes larger than 2.5 times the rms value for the ion saturation current (full blue line), the synthetic data (dotted black line) and the best fit of a double-exponential pulse shape to the measurement data (dashed green line).}
\label{fig:jcav}
\end{figure}

In \Figref{fig:jcav} the conditionally averaged wave-form for the ion saturation current is presented for peak fluctuation amplitudes larger than 2.5 times the root mean square value and a conditional window duration of $200\,\mu$s. This resulted in a total of 7471 non-overlapping events for this long data time series. The saturation current wave-form has an asymmetric shape with a fast rise and slower decay, as is also apparent in the raw data presented in \Figref{fig:jraw}. The average wave-form is well described by a double-exponential pulse shape with a rise time of $11\,\mu$s and fall time of $19\,\mu$s, giving a duration time of $30\,\mu$s, in agreement with the correlation analysis presented above. While the underlying pulses for the synthetic data have a sharp rise, the conditionally averaged wave-form has a finite rise time due to pulse overlap. The difference in the shape of the conditional wave-forms for the measurement and synthetic data is therefore as expected. Note that the peak amplitudes are in perfect agreement.

\begin{figure}
\includegraphics[width=7.5cm]{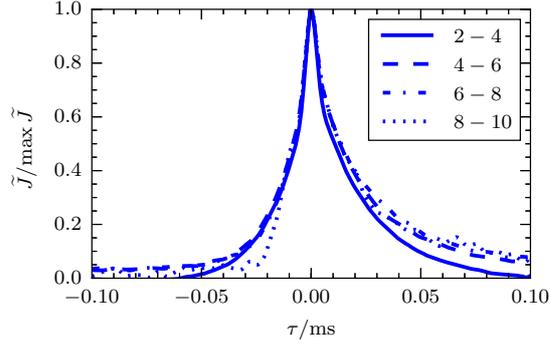}
\caption{Conditionally averaged burst wave-forms for the ion saturation current signal with peak amplitudes in units of the rms value given by the range indicated in the legend.}
\label{fig:jcav-ampl}
\end{figure}

Restricting the peak amplitude of conditional events in the ion saturation current signal to be within a range of 2--4, 4--6, 6--8 and 8--10 times the rms value, the appropriately scaled conditional wave-forms, shown in \Figref{fig:jcav-ampl}, reveal that the average burst shapes and durations do not depend on the burst amplitude and are again well described by a double-exponential wave-form. This gives further support for the assumptions underlying the stochastic model presented in \Secref{sec:model}.

\begin{figure}
\includegraphics[width=7.5cm]{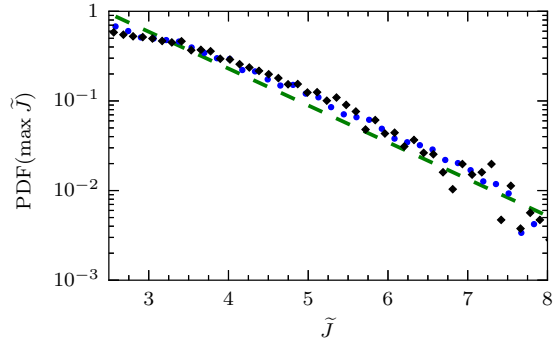}
\caption{Probability distribution function for burst amplitudes with peak values larger than 2.5 times the rms level for the ion saturation current (blue circles), the synthetic data (black diamonds), and an exponential fit to the measurement data (dashed green line).}
\label{fig:jamp}
\end{figure}

For conditional burst events, the peak amplitudes after the signal crosses a certain threshold level are also recorded. \Figref{fig:jamp} shows the distribution of these peak amplitudes for ion saturation current and synthetic data fluctuations larger than 2.5 times the rms level. This is clearly well described by a truncated exponential distribution, as might be expected from the exponential tail in the distribution function for the full signal presented in \Figref{fig:jpdf}. The mean value of the fitted exponential distribution is $3.6$, consistent with the peak amplitude of the conditionally averaged ion saturation current wave-form shown in \Figref{fig:jcav}. There is excellent agreement for the amplitude distribution between the measurement and synthetic data.

\begin{figure}
\includegraphics[width=7.5cm]{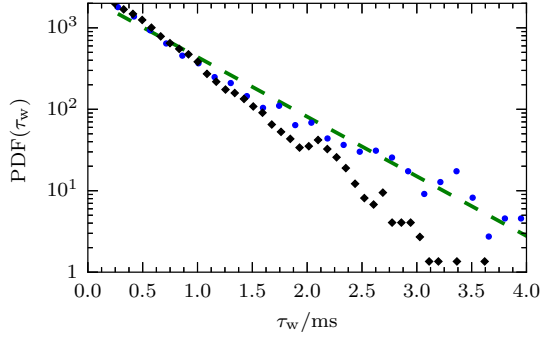}
\caption{Probability distribution function for waiting times between large-amplitude events with peak values larger than 2.5 times the rms level for the ion saturation current (blue circles), the synthetic data (black diamonds) and an exponential fit to the measurement data (dashed green line).}
\label{fig:wait}
\end{figure}

From the occurrence times of large-amplitude events in the ion saturation current signal, the waiting times between them is also calculated. As shown in \Figref{fig:wait}, for peak amplitudes larger than 2.5 times the rms value, the waiting time distribution is well described by an exponential function over three orders of magnitude on the ordinate. The mean value of the waiting times based on this fit is $0.8\,$ms. Such an exponential distribution of waiting times is in accordance with a Poisson process, suggesting that large-amplitude fluctuations in the far-\sol\ are uncorrelated. A similar analysis of the synthetic data also reveals exponentially distributed waiting times, but the average waiting time is slightly shorter than for the measurement data. The reason for this has yet to be clarified.

\begin{figure}
\includegraphics[width=7.5cm]{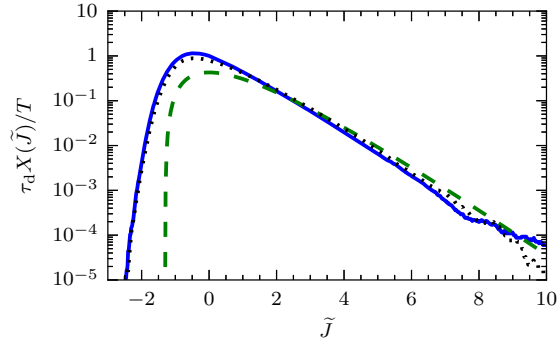}
\caption{Rate of level crossings for the ion saturation current (full blue line), synthetic data (dotted black line) and predictions from the stochastic model without additive noise (dashed green line).}
\label{fig:xing}
\end{figure}

For a time series of duration $T$, the number of up-crossings over the level $\wt{J}$ is denoted by $X(\wt{J})$. The normalised rate of such level crossings is presented in \Figref{fig:xing} for both the measurement and synthetic data time series. This is compared to an analytical prediction for the stochastic model described by \Eqref{fpp} \cite{theodorsen-php,garcia-php}. The number of up-crossings of the $2.5\Jrms$-level is 18298 for the ion saturation current time series. As expected, the rate of level crossings is largest around the mean value of the signal. The analytical model under-estimates the rate of level crossings for low threshold levels, which is obviously due to the additional noise in the measurement and synthetic data time series. However, the tail behaviour for large threshold levels compares favourably with the analytical expression. For all threshold levels, there is excellent agreement between the measurements and the synthetic data.

\begin{figure}
\includegraphics[width=7.5cm]{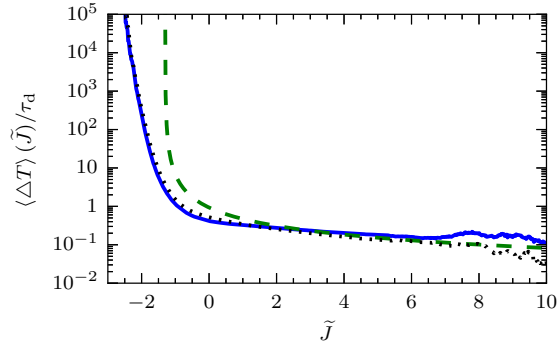}
\caption{Average excess times for the ion saturation current (full blue line), synthetic data (dotted black line) and predictions from the stochastic model without additive noise (dashed green line).}
\label{fig:tave}
\end{figure}

For the stationary stochastic process described by \Eqref{fpp} it is possible to calculate analytically both the number of up-crossings and the total time spent above a given threshold level, the latter given by the complementary cumulative distribution function. The ratio of these gives the average duration $\langle\triangle T\rangle$ of time intervals spent above the threshold level \cite{theodorsen-php,garcia-php}. In \Figref{fig:tave} this theoretical prediction is compared to direct computations of the average excess times for the measurement and synthetic data time series. Since both the distribution function and level crossing rate are well described by the model realisation, the excellent agreement between measurement and synthetic data in \Figref{fig:tave} comes as no surprise. As for the level crossing rate, the analytical model without additive noise fails to accurately describe average excess times for low threshold levels. For large threshold levels, the average duration of excess times is slightly smaller than the pulse duration $\taud$ and decreases gradually with the threshold level.

\section{Discussion and conclusions}\label{sec:discuss}

Langmuir probe measurements at the outboard mid-plane region of \kstar\ have revealed results that are consistent with observations in many other devices. The \sol\ generally exhibits a two-layer structure: a near-\sol\ with a steep profile and moderate fluctuation level near the separatrix, and a flatter profile with larger fluctuations outside this in the so-called far-\sol. As the line-averaged plasma density increases, the profile scale length in the far-\sol\ increases and the break point between the near- and far-\sol\ regions moves radially inwards. This substantially enhances plasma interactions with the main chamber walls.

The large profile scale length and fluctuation level in the far-\sol\ region is generally attributed to the radial motion of blob-like plasma filaments. The stochastic model outlined in \Secref{sec:model} predicts an exponential radial profile for a super-position of propagating pulses with constant size and velocity \cite{garcia-php},
\begin{equation}
\Phiave(r) = \frac{\taud}{\tauw}\ave{A}\exp\left( - \frac{r}{V_\perp\tau_\parallel}\right) ,
\end{equation}
where the parallel transit time is estimated by the ratio of the magnetic connection length $L_\parallel$ and the sound speed $\Cs$, $\tau_\parallel=L_\parallel/\Cs$. Since the connection length and electron temperature typically remain constant in the far-\sol, this suggests that the increase in the profile scale length is due to faster blob-like plasma filaments. However, the average density in the \sol\ may also increase due to higher pulse amplitudes $\ave{A}$ and stronger pulse overlap.

The far-\sol\ in \kstar\ is characterised by large relative fluctuation levels and positively skewed and flattened fluctuations. Similar to observations on several other tokamaks, these moments vary little with radial position and line-averaged density \cite{labombard1,garcia-tcv-jnm,garcia-tcv-nf,rudakov-nf}. This suggests that while the fluctuations are strongly intermittent, they have universal properties. These properties have been explored by a novel experiment on \kstar\ in which the probe was dwelled in the far-\sol\ during an entire discharge in order to obtain a time series of the ion saturation current under stationary plasma conditions of unprecedented duration. It is found that large-amplitude bursts on average have an exponential wave-form with exponentially distributed burst amplitudes and waiting times. Moreover, the burst shape and duration does not depend on the burst amplitude, similar to previous results from Alcator C-Mod and TCV \cite{garcia-acm-jnm,garcia-acm-php,garcia-tcv-nfl,theodorsen-tcv-ppcf,kube-acm-ppcf}.

These are exactly the assumptions underlying a recently proposed stochastic model for the intermittent plasma fluctuations described as a super-position of uncorrelated pulses \cite{garcia-prl,garcia-php,theodorsen-php}. Consistent with predictions of this model, the auto-correlation function for the ion saturation current time series is found to be exponential, the \pdf\ is given by a Gamma distribution and there is a parabolic relation between the skewness and flatness moments. By adding random noise to this process, an identical analysis of a model simulation and the measured ion saturation current are in excellent agreement, demonstrating that the stochastic process reproduces all the salient statistical properties of the plasma fluctuations.

Based on the stochastic model, novel predictions have been given for the rate of level crossings and the average duration of time intervals spent above a specified threshold level \cite{garcia-php,theodorsen-php}. By adding random noise, a realisation of the process have been shown to give predictions of these quantities that are in excellent agreement with the experimental measurements. Provided the fluctuation statistics have universal properties, an experimental determination of the correlation time and the lowest order statistical moments are thus sufficient in order to predict the distribution of fluctuation amplitudes, level crossing rates and excess times in the vicinity of the main chamber walls. These quantities are particularly relevant for plasma--surface interaction  processes such as sputtering and melting, which are threshold phenomena.

\section*{Acknowledgements}

This research was partially supported by Ministry of Science, ICT, and Future Planning under KSTAR project and was partly supported by National Research Council of Science and Technology (NST) under the international collaboration and research in Asian countries, PG-1314. The views and opinions expressed herein do not necessarily reflect those of the ITER Organization. ITER is the Nuclear Facility INB-174. OEG, RK and AT were supported with financial subvention from the Research Council of Norway under grant 240510/F20.

\end{document}